\documentclass{webofc}
\usepackage[utf8]{inputenc}

\title{Evaluation of research publications and publication channels in astronomy and astrophysics}
\author{\firstname{Eva} \lastname{Isaksson}\inst{1}\fnsep\thanks{\email{Eva.Isaksson@helsinki.fi}} \and 
\firstname{Henrik} \lastname{Vesterinen}\inst{2}}
\institute{Helsinki University Library, University of Helsinki, Finland 
           \and University of Tampere, Finland}
\titlerunning{Evaluation of research publications}
\authorrunning{Isaksson & Vesterinen}
\date{June 2017}

\abstract{The astronomy community usually turns to the Astrophysics Data System for bibliometrics. When the context is cross-disciplinary, commercial products like Web of Science and Scopus are used along with related analytics tools instead. The results are often tainted by inherent problems in the chosen classification system. A review of the most common challenges and pitfalls is given. 

Commercial altmetrics products could be added to the evaluation toolbox in the near future despite the fact that they are best suited for promotion instead of evaluation.

Norway, Denmark, and Finland have created journal and publisher ranking systems that are used in national funding models.  Differences in how astronomy journals are weighed in these systems night be related to the volume of papers published on a national level.}

\begin{document}

\maketitle

\section{Introduction}

The growth of evaluation culture in the 2000s has produced a number of tools and classifications to automate the assessment of university research outputs. While the Astrophysics Data System remains the preferred bibliometric tool for the astronomy community, it is not applicable in evaluations that span a wide range of research areas. University rankings and other evaluation processes use bibliometric data from either Web of Science or Scopus, while universities increasingly choose to purchase related do-it-yourself analytics tools.
 
In this paper the focus is on mainstream bibliometrics, using commercial bibliometrics products. We have checked how well they can handle evaluation of astronomy and astrophysics publications.

We will cover Web of Science, Scopus, and the analytics toos they use (InCites and SciVal). Our examples are mostly from Nordic countries. We also checked altmetrics, and the so called Nordic model of ranking journals. The time frame used is 2000-2015.

A key question when we turn to other databases and tools is: How do we define astronomy, if we want the results to be useful and realistic. The astronomy community has a pretty good idea as to which journals cover astronomy. That is not so clear when using commercial databases.

The usual way is to pick a research category defined by either Web of Science or Scopus.  Both of these have their merits and problems.

\section{Web of Science and InCites: the HEP dilemma}

Web of Science by Clarivate Analytics (formerly Thomson \& Reuters Research Intelligence) is the preferred database used in Nordic comparisons of institutes \cite{Piro}. It uses about 250 research categories. Each journal is listed in one or more categories. There were 63 journals in the astronomy \& astrophysics category for 2016. If a journal is listed both in physics, particles and fields (PPF) and in astronomy \& astrophysics (AA) categories, all of its articles are included in both, no matter what each article actually contains. Space and planetary physics journals tend to get an additional geosciences (GEO) related category.

To make the process more straightforward, we only looked at 21 top journals, where the Nordic countries (Sweden, Denmark, Finland, Norway) publish 95\% of their astronomy papers. Next, we divided the journals into three groups according to which categories they belong:

\begin{itemize}
\small
    \item \textbf{AA category only:} \textit{Astronomy \&Astrophysics, Astronomical Journal, Astronomische Nachrichten, The Astrophysical Journal, The Astrophysical Journal Letters, The Astrophysical Journal Supplement, Astrophysics and Space Science, Icarus, Monthly Notices of the RAS, Planetary and Space Science,  Solar Physics, Space Science Review }  (12 journals)
    \item \textbf{AA \& PPF categories:} \textit{Physical Review D, Physics Letters B, Journal of Cosmology and Astroparticle Physics, Classical \& Quantum Gravity, Astroparticle Physics}  (5 journals)
    \item \textbf{AA \& GEO categories:} \textit{Journal of Geophysical Research: Space Physics, Annales Geophysica, Advances in Space Research, Radio Science}  (4 journals)
\end{itemize}

Looking at the number publications, it turns out that there are some differences between Nordic countries regarding these groups. The AA group was strongest (58\% of the total Nordic output), except for Norway, where the PPF was on top. After \textit{Physical Review D} was added to the category in 2005, the share of the PPF group has become significant (28\% of the output).  The GEO group was the smallest everywhere (14\%), and it was particularly small in Denmark. What are the effects on such differences? 

To see the impact of a set of research publications, one usually normalizes the citations so to reflect the typical number of citations for a particular research category in a chosen time frame. The world average for each category is set to one (1). Impacts higher than 1 are said to be above world average for a particular category.

\begin{figure}[h]
\centering
\sidecaption
\includegraphics[width=7cm, clip]{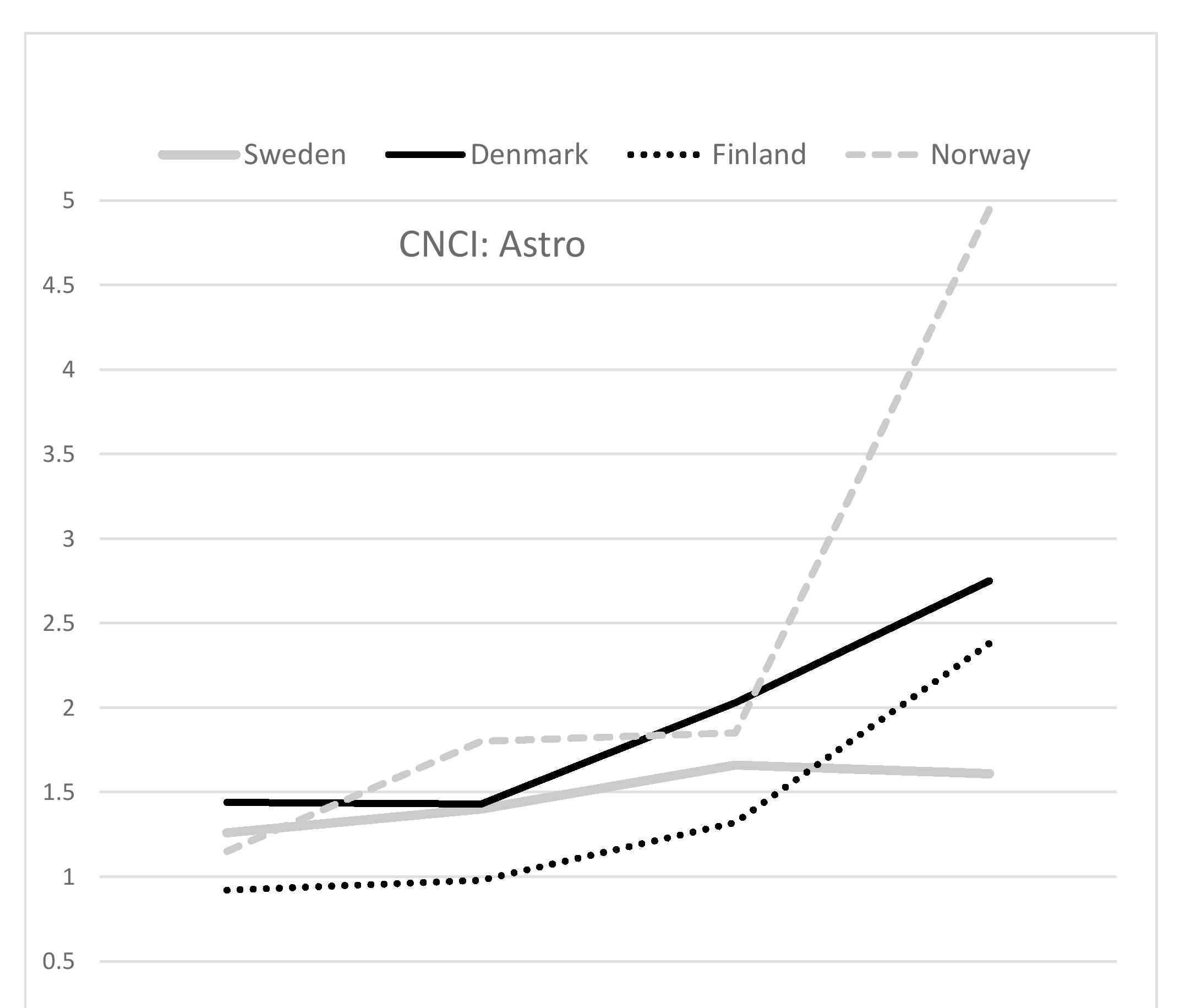}
\caption{Category Normalized Citation Impact (CNCI) for Nordic Countries for the astronomy, physics and geosciences journal subgroups of the Web of Science Astronomy \& Astrophysics research category, 2000--2015. World average is set to 1.\newline \textit{\scriptsize{© Copyright CLARIVATE ANALYTICS ® 2017.  All rights reserved. InCites dataset updated  2017-05-13. Includes Web of Science content indexed through 2017-03-31.}} }
\includegraphics[width=7cm,clip]{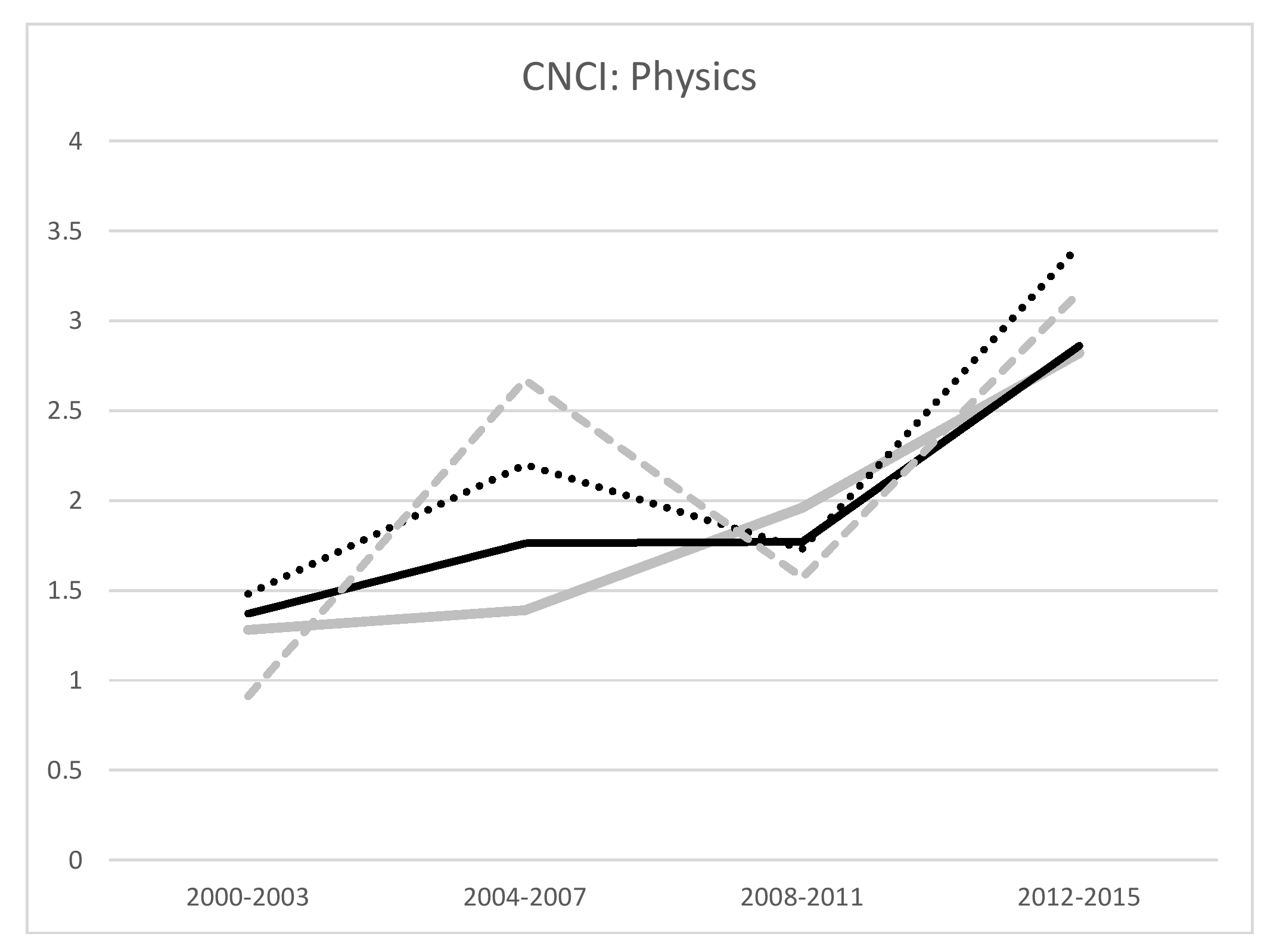}
\includegraphics[width=7cm,clip]{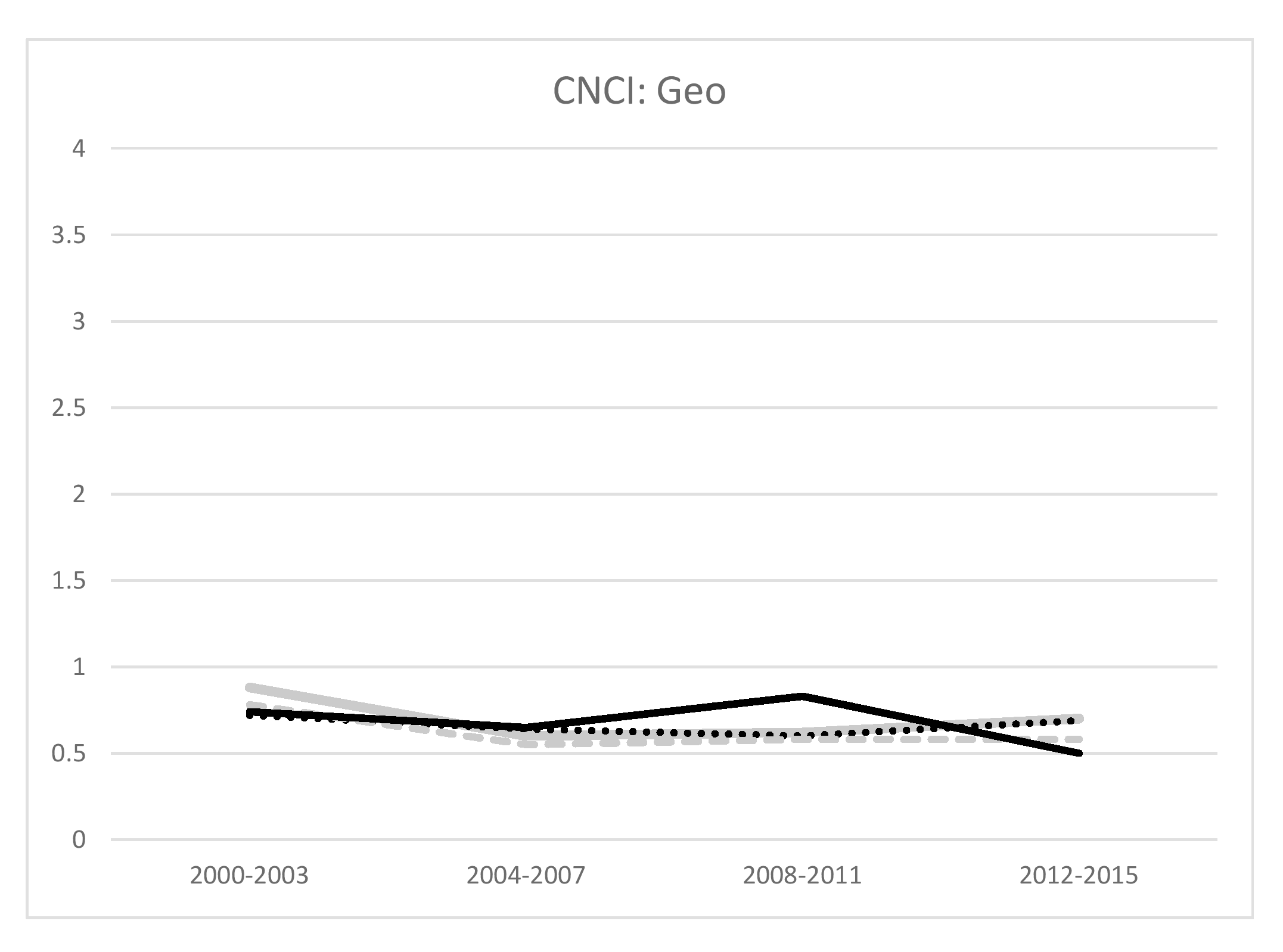}
\label{fig-1}  
\end{figure}

Figure~\ref{fig-1} shows the category normalized results for Sweden, Denmark, Finland and Norway. The starting point is year 2000 and the publications are smoothed into four year time frames.

The results for the AA and physics group of journals differ considerably from the geosciences group. Astronomy and physics journal groups are trending upward, well above the world average. This is probably caused by participation in the very big, high impact collaborations like Planck. In comparison, the trend for the geosciences journal group does not look too good. That does not necessarily mean that the astronomy published in these journals has very little impact. Instead, they are being compared to journals with a different citation rate.

One should keep in mind that we only see some general trends, as InCites does not fractionalize the numbers to reflect the varying number of countries, institutions, groups or authors. 

\section{Trend Visualizations with VOSviewer Maps}

Is there a way to check what these journals really contain?
Are the journal groups really different from each other? One way to look inside is to use term maps. With VOSviewer,\footnote{http://www.vosviewer.com} you can map terms from titles and abstracts of publications.

\begin{figure}[h!]
\centering
\includegraphics[width=1\textwidth]{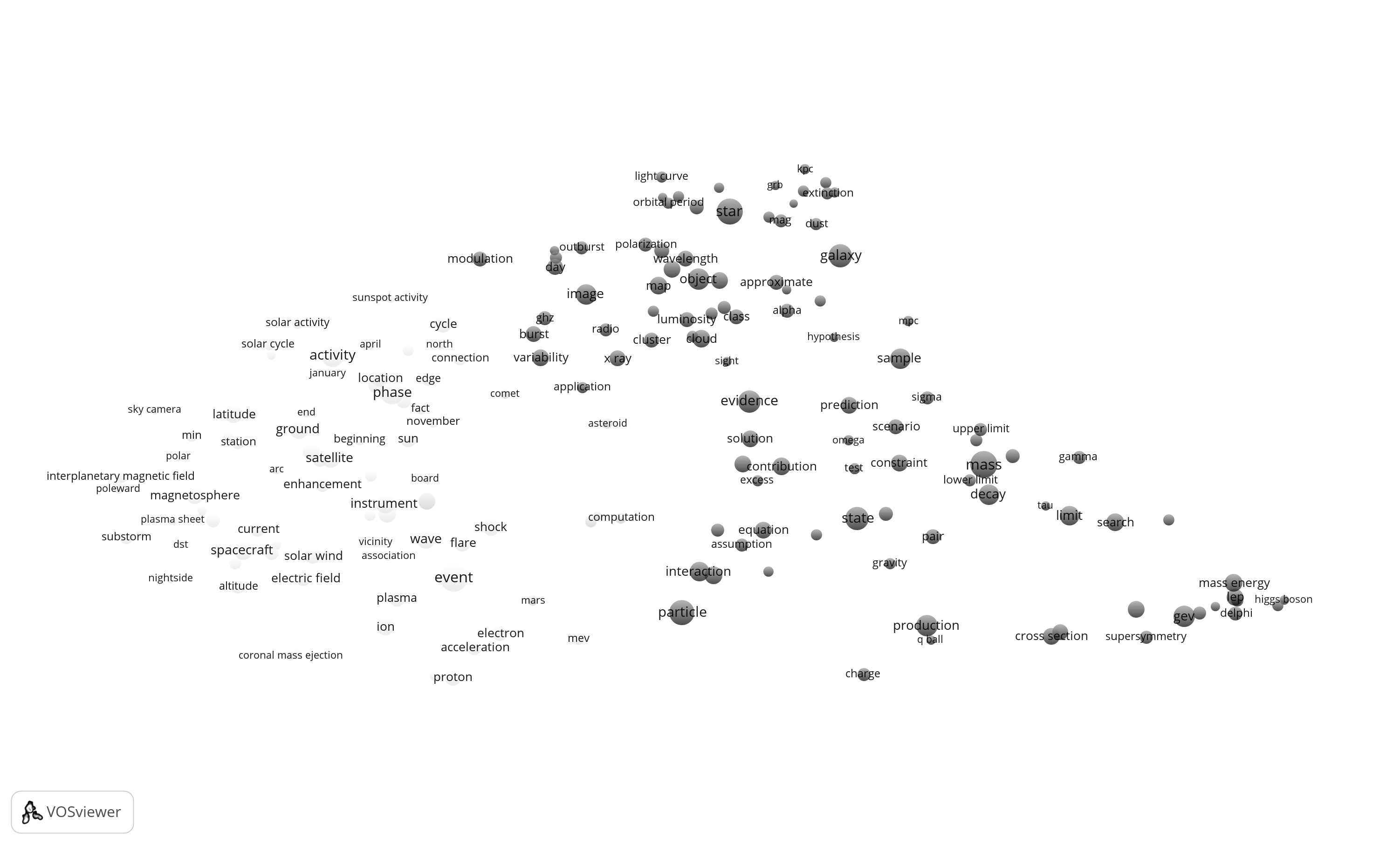}
\includegraphics[width=1\textwidth]{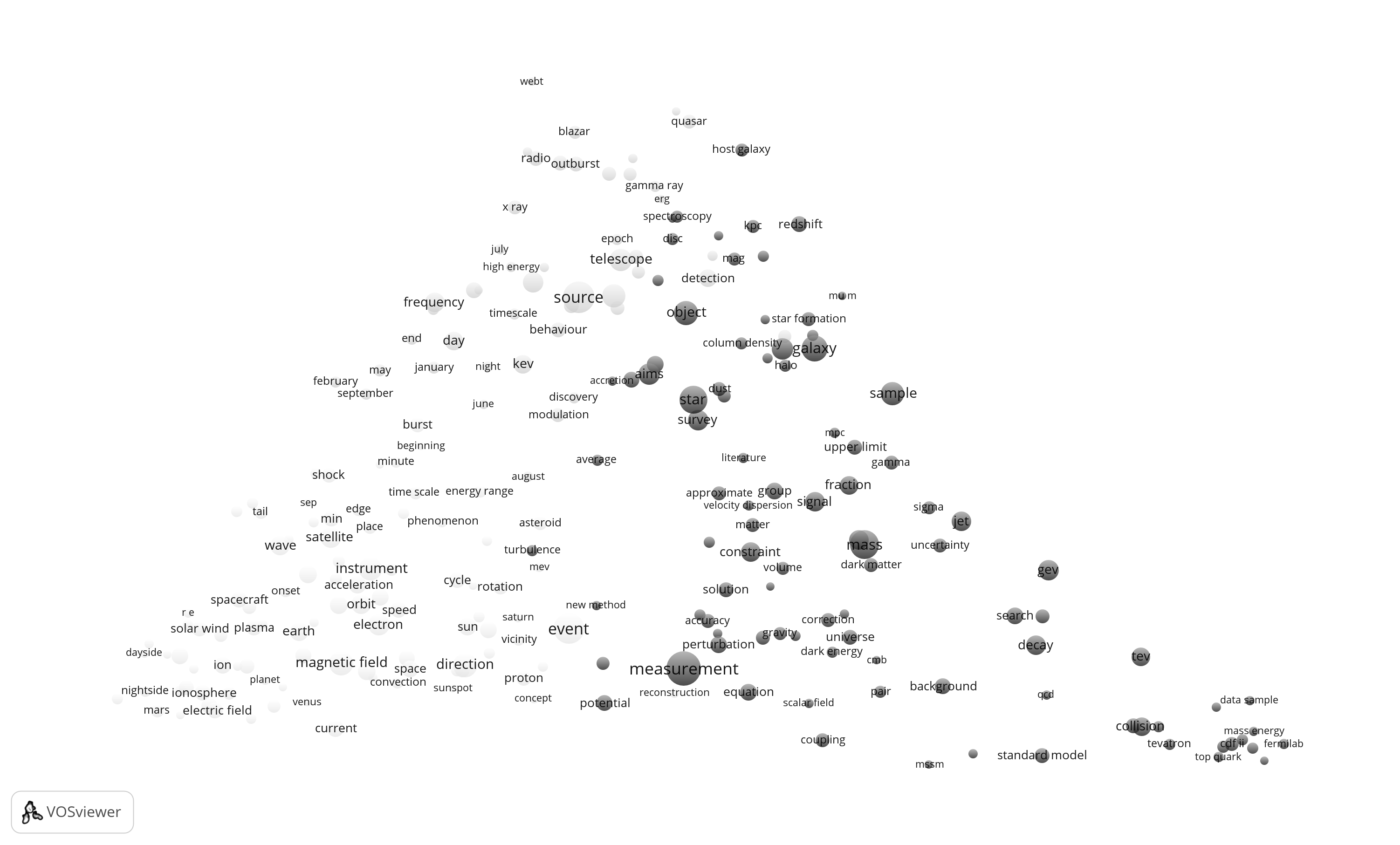}
\caption{Term maps for years 2000-2003 (top) and 2004-2007 (bottom). More or less distinct clusters of astronomy, planetary sciences, and high energy physics are seen here. Astronomy can be seen in the top portions of the figures, planetary sciences on the left, and HEP on the right.}
\label{wos2000-2007}
\end{figure}

\begin{figure}[h!]
\includegraphics[width=1\textwidth]{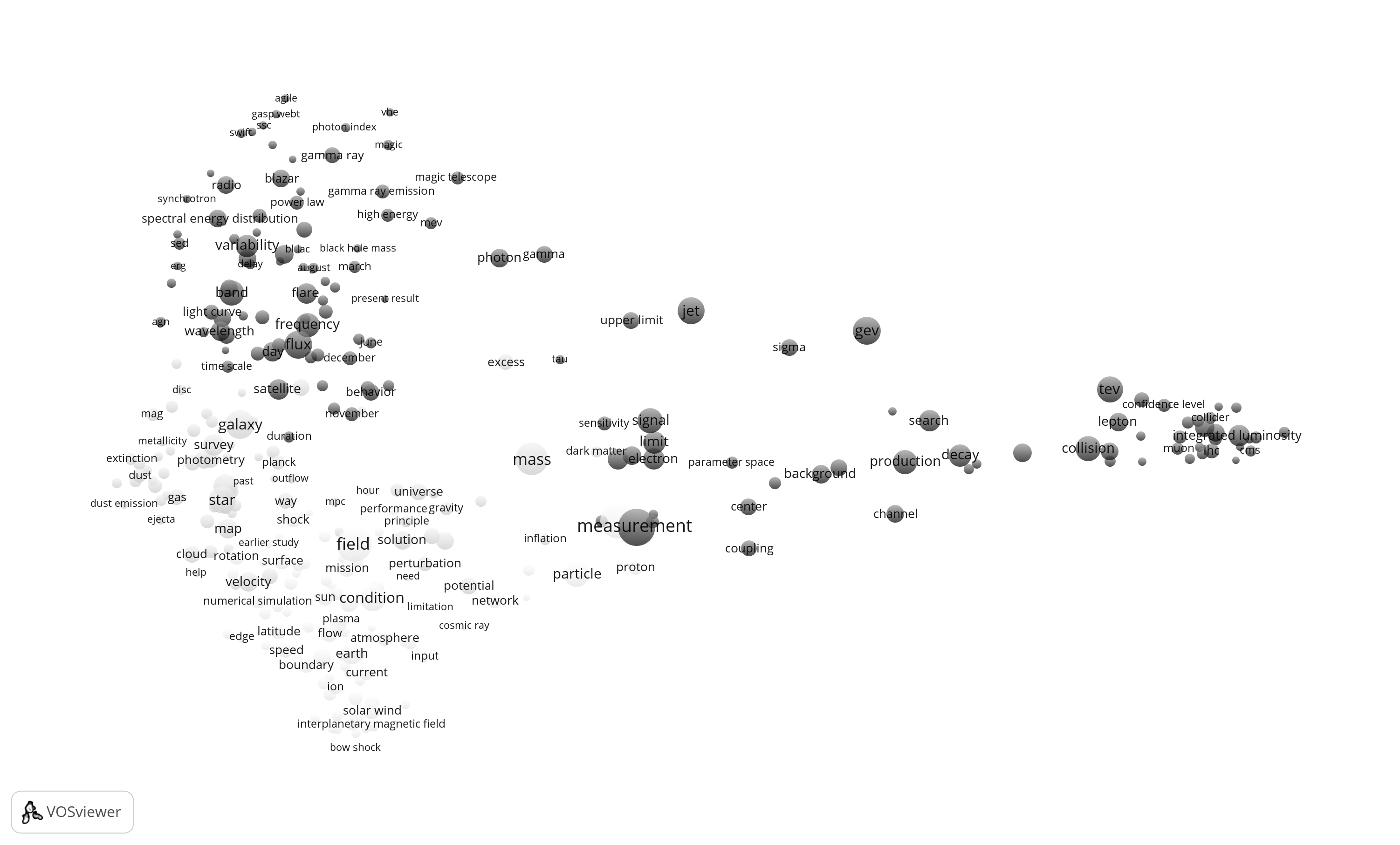}
\includegraphics[width=1\textwidth]{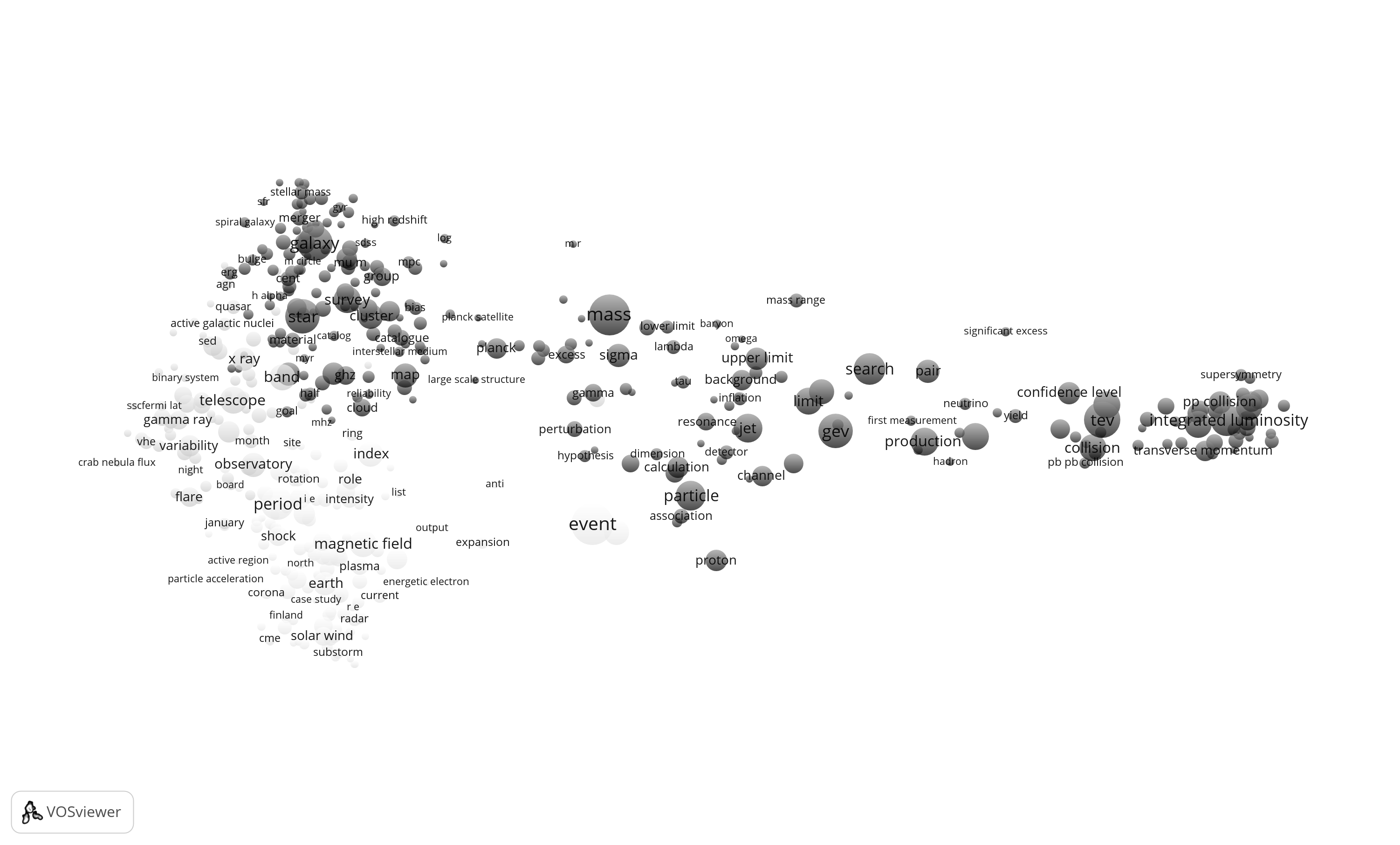}
\caption{Term maps for years 2008-2011 (top) and 2011-2015 (bottom). Clusters of astronomy, planetary sciences, and high energy physics are clearly seen here. Of special interest is the apparent drifting of HEP away from astronomy.}
\label{wos2008-2015}
\end{figure}

Figures \ref{wos2000-2007} and \ref{wos2008-2015} contain term maps (clusters) for the selected journals categorized as astronomy by Web of Science. The underlying data consist of the titles and abstracts of all the papers published in the journals in Finland in the time period 2000-2015.

The term maps in Figure \ref{wos2000-2007} are from the start of the time period we are looking at. We can see distinct clusters for the astronomy, PPF and geosciences journal terms, but they are not that much apart. Figure \ref{wos2008-2015} shows the same group of journals but many years later. The astronomy and geophysics groups are now closer together. The PPF content has drifted into a direction of its own.

\clearpage
The above results (CNCI and VOSviewer) show that the presence of two strong High Energy Physics (HEP) journals strongly affects the content and impact for the astronomy \& astrophysics research category in the Web of Science. If we look at the worldwide results for this category, we see that \textit{Physical Review D} tops the list. In Nordic countries, it is near the top, while the number one journal is still \textit{Astronomy and Astrophysics}. This strong HEP presence means that astronomy evaluations are in fact astronomy and HEP evaluations when you use Web of Science categories. 

Astronomy and HEP are obviously linked by subjects like dark matter and dark energy (see e.g., essay by Kolb \cite{Kolb}). But how to make a clear distinction between these two, and is it even possible? The dividing line is not linearly positioned between the fields. Astronomy has also been evolving towards large, HEP-style collaborations. That too will necessarily show up in bibliometric analysis. 

In the period we looked at, experimental HEP publications have followed the pattern discussed by Pritychenko \cite{Pritychenko} for nuclear physics. There are fewer small facilities, and research is increasingly done with big instruments and by large collaborations connected by data-intensive networks. 

The geosciences oriented journals suffer in comparison. One is tempted to ask whether you can rise above the world average if you are competing in the same category with LHC and the Higgs boson.

\section{Scopus and SciVal: Better without HEP?}

Scopus is a part of Elsevier and has established itself as the other big bibliometric service. In many respects, it is similar to Web of Science, with an analysis tool of its own, SciVal. There is a journal classification system that assigns one or more categories to each journal. The interesting thing is that SciVal has a more detailed classification than Scopus. On the other hand, the impact is fractionalized between categories, which tends to even out the differences.

For astronomy, there are two main categories: Astronomy \& Astrophysics (A\&A), and Planetary \& Space Science (P\&SS). Most astronomy journals belong to both, which means that these journals have to cut their results in half between both. For the top journals in the Nordic countries, the main differences for these two groups seemed to be a lone non-P\&SS journal, \textit{Journal of Astronomy and Astroparticle Physics}, and a number of geosciences oriented journals that were not included in the A\&A group. 

\begin{figure}[h]
\centering
\sidecaption
\includegraphics[width=11cm,clip]{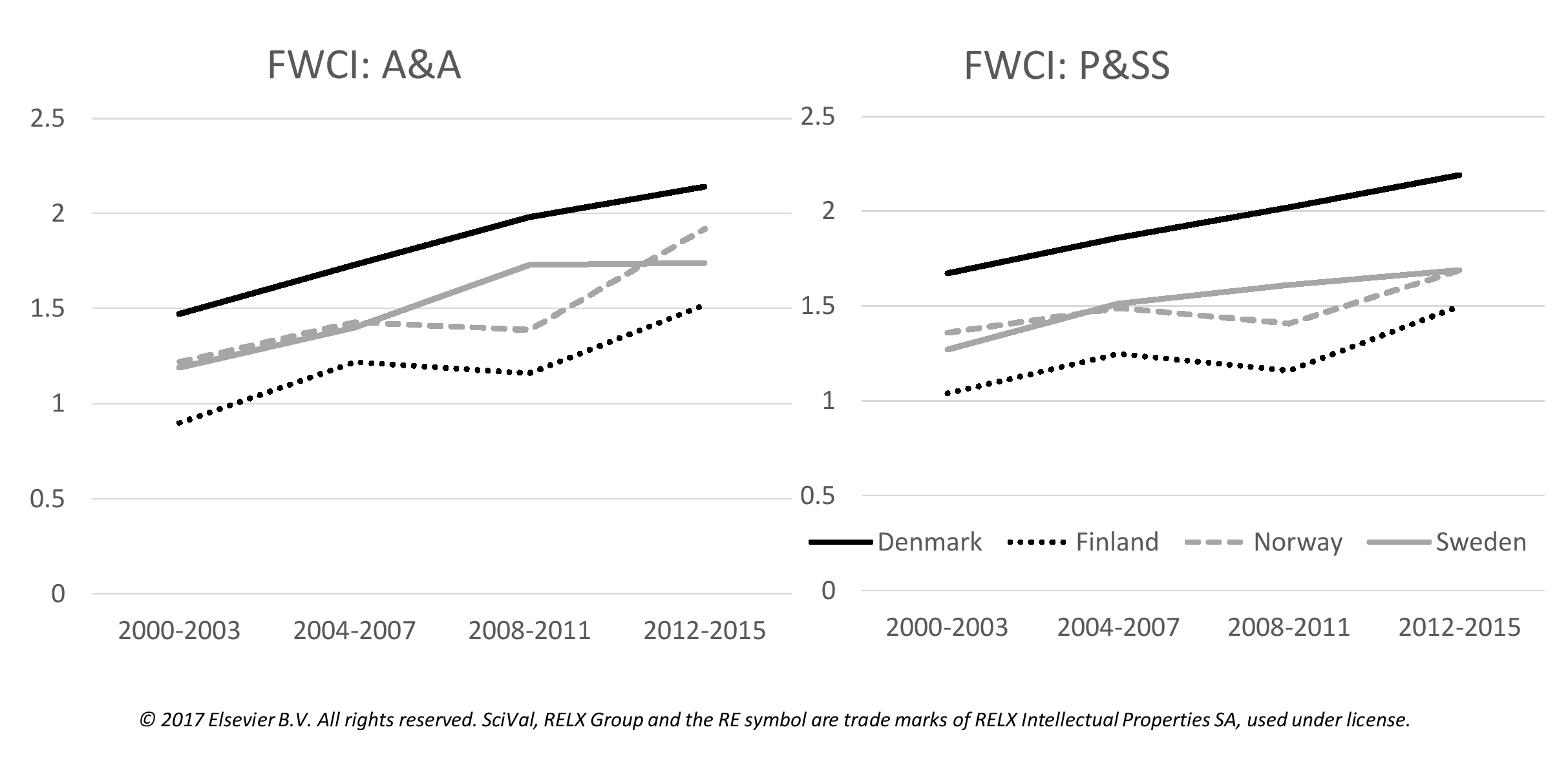}
\caption{Field Weighed Citation Impact for Nordic Countries for SciVal categories Astronomy and Astrophysics (A\&A) and Planetary \& Space Sciences (P\&SS).}
\label{fig-3}       
\end{figure}

SciVal has a citation normalization indicator of its own,
called Field Weighed Citation Impact (FWCI). Figure~\ref{fig-3} shows the trends for Nordic countries. These results probably look good to most people. There are no HEP journals included, so administrators can see that their institution is above world average, and going upwards, no matter whether you look at the A\&A or the P\&SS category.

The halving of the results into two categories produces some notably distortions. Denmark is stronger in the A\&A category, compared to the P\&SS category. However, as astronomy journals tend to be found in both categories, it looks like the A\&A results for Denmark are lower and the P\&SS ones higher that they actually are.

\section{Altmetrics: Mentions versus Impact}

In the 2010s, we have seen the rise of altmetrics or alternative metrics that aims to catch the immediate, societal impact of research outputs by counting mentions from channels such as social media, blogs and news. Can we consider it as an additional tool in our evaluation toolbox?

What exactly are we measuring when we are counting mentions? What does a metric like mentions per output mean for a set of publications? To see how mentions are related to normalized impact, we took the Web of Science astronomy \& astrophysics research area and chose the journal groups listed earlier in this paper. Instead of Nordic countries, we used
the total worldwide output for these journals. Mentions were harvested from Altmetric Explorer for 2010--2016 (older results are not comprehensive enough). Impact was found from InCites for the same period.

\begin{figure}[h]
\centering
\sidecaption
\includegraphics[width=10cm,clip]{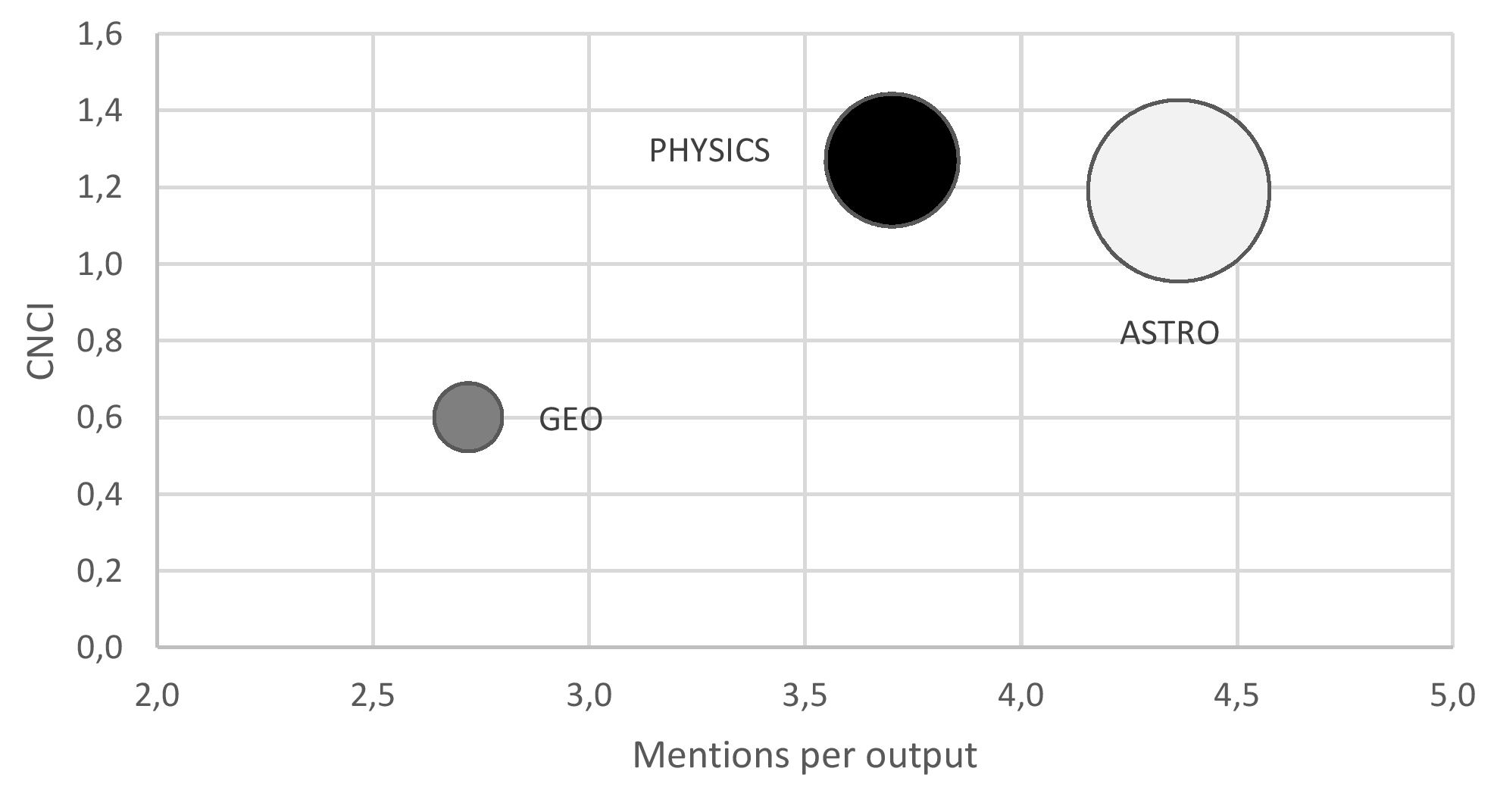}
\caption{Category Normalized Citation Impact (CNCI) versus mentions per output for the astronomy, physics and geosciences journal subgroups of the Web of Science Astronomy \& Astrophysics research category, 2010--2016. World average is set to 1.}
\label{fig-4}      
\end{figure}

The result is shown in Figure~\ref{fig-4}. The astronomy group of journals has most mentions (4.4 per output) compared with particles and fields group of HEP journals (3.7) and geosciences journals (2.7). In particular, the astronomy journal group excels in wikipedia and news mentions. As for CNCI values, the geosciences journal group is clearly below world average with 0.6.  The HEP journal group scores 1.3 and the astronomy group 1.2.

What do the numbers mean? It is clear that the geosciences group of journals are getting less attention in terms of mentions. Are they any worse for that? These results tell us little about the mechanisms behind the numbers, and consequently, they are not suited for evaluation.

\section{Journal levels}

In the Nordic countries (except for Sweden), national journal ranking systems have been introduced mainly as a tool for funding university research.\footnote{One can use the search tool at the Finnish portal for looking at journal levels in Nordic countries:\\ https://www.tsv.fi/julkaisufoorumi/haku.php?lang=en} In this system, a certain percent of the funding is distributed to the universities according to where and how much they publish. Journals and publishers are assigned levels determined by expert panels. Also those publication channels not found in Web of Science or Scopus (whether representing social sciences and humanities which have a low coverage in those databases, or published in national languages) are assigned a level. 

Levels do not reflect the impact of research directly, as they do not measure the impact of individual papers. Also, they need to be balanced between disciplines, so that only a certain percentage of journals are assigned to a higher level for a given discipline. Levels often do not reflect the real quality of a journal.

\begin{table}[h!]
\footnotesize
\centering
\begin{tabular}{ |l|c|c|c| } 
\hline
Journal & Finland & Norway & Denmark \\
\hline
A\&A	        & 3	& 2	& 2 \\
A\&A Rev.	        & 3	& 1	& 2 \\
ApJ	        & 2	& 2	& 2 \\
ApJ Lett.   & 1	& 2	& 1 \\
ApJS        & 1	& 2 & 1 \\
JCAP             & 1 & 1 & 2 \\
JGR: Space Phys. & 3 & 1 & 2 \\
MNRAS       & 1	& 1	& 2 \\
Phys. Rev. D     & 2 & 1 & 1 \\
Phys. Lett. B    & 2 & 1 & 1 \\
\hline
\end{tabular}
\caption{Journal levels (June 2017) are generally not identical for Finland, Norway and Denmark.}
\end{table}

Journal levels are occasionally used for evaluation of units and individual researchers, against official recommendations. In Finland, the Ministry of Education compared both impact and publication points from journal levels for Finnish universities\cite{OKM}. According to the analysis, astronomy publications by University of Helsinki had bigger impact compared to University of Turku. In publication points, the outcome was reversed for so-called productivity. In a situation like this, you can trust either Web of Science or journal levels. The latter tell very little about actual impact of research.

Table 1 lists some interesting diffences between Nordic countries. The only journal all three systems agree on is \textit{The Astrophysical Journal}. For other journals, there is a rough rule of thumb: the more you publish in a journal, the more likely your national panels are to choose a higher level. This seems to be true for Finland and Denmark. These different levels for the same journals could lead to some variation as how  astronomy output is rewarded in terms of  national research funding.

\section{Conclusion}

Research evaluations are usually built on comparisons of units. For an international effort like astronomy, the results are better when one participates in high profile international collaborations. Certain journal titles that publish such research will influence the expected impact of whole research areas. 
Evaluations that span several disciplines cannot use the Astrohysics Data System, as there are no normalization tools for the comparisons that are usually needed. The next best choice could be Web of Science with InCites, closely followed by Scopus with SciVal. Both of these have their own problems. Web of Science includes two extremely strong HEP journals in the Astronomy and Astrophysics Research Area. As a result, HEP results tend to overshadow some of the astronomy and astrophysics. There is also a geosciences (space physics, planetary science) related group of journals that suffer even more from the HEP presence. 

In Scopus and SciVal, the Planetary Sciences have a category of their own, without any HEP journal complications. However, the fractionalizing of results between categories distorts the impact for most astronomy journals. 

As for altmetrics, it is certain that astronomy is one of the disciplines with a significant number of mentions. There is a very long way from that to meaningful metrics. Instead of evaluations, altmetrics is better suited for promotion and visibility of research.

Of all approaches to evaluation, national journal ranking systems are the most problematic, as they were developed to serve as tools for university funding.

\end{document}